\documentclass{vgtc}                          

\ifpdf
  \pdfoutput=1\relax                   
  \usepackage{graphicx}                
  \DeclareGraphicsExtensions{.pdf,.png,.jpg,.jpeg} 
\else
  \usepackage{graphicx}                
  \DeclareGraphicsExtensions{.eps}     
\fi%

\graphicspath{{figures/}{pictures/}{images/}{./}} 

\usepackage{microtype}                 
\PassOptionsToPackage{warn}{textcomp}  
\usepackage{textcomp}                  
\usepackage{mathptmx}                  
\usepackage{times}                     
\usepackage{cite}                      
\usepackage{tabu}                      
\usepackage{booktabs}                  
\usepackage{algpseudocode}
\usepackage{amsmath,amssymb}
\usepackage{algorithm}
\usepackage[caption=false]{subfig}

\title{Interactive graph query language for multidimensional data in Collaboration Spotting visual analytics framework}

\author{Adam Agocs\thanks{e-mail: Adam.Agocs@cern.ch}\\ %
        \scriptsize CERN, CH-1211 Geneva 23, Geneva, Switzerland %
\and Dimitrios Dardanis\thanks{e-mail: Dimitrios.Dardanis@cern.ch}\\ %
     \scriptsize CERN, CH-1211 Geneva 23, Geneva, Switzerland%
\and Jean-Marie Le Goff\thanks{e-mail: Jean-Marie.Le.Goff@cern.ch}\\ %
     \scriptsize CERN, CH-1211 Geneva 23, Geneva, Switzerland
\and Dimitrios Proios\thanks{e-mail: Dimitrios.Proios@cern.ch}\\
	\scriptsize CERN, CH-1211 Geneva 23, Geneva, Switzerland
     }

\abstract{Human reasoning in visual analytics of data networks relies mainly on the quality of visual perception and the capability of interactively exploring the data from different facets. Visual quality strongly depends on networks' size and dimensional complexity while network exploration capability on the intuitiveness and expressiveness of user frontends.
 
The approach taken in this paper aims at addressing the above by decomposing data networks into multiple networks of smaller dimensions and building an interactive graph query language that supports full navigation across the sub-networks. Within sub-networks of reduced dimensionality, structural abstraction and semantic techniques can then be used to enhance visual perception further.
} 

\CCScatlist{ 
\CCScat{Information Systems}{H.2.3}{Language}{Query language}
\CCScat{Image Processing and Computer Vision}{I.4.10}{Image Representation}{Multidimensional}
}

\begin{document}


\newtheorem{theorem}{Theorem}
\newtheorem{definition}{Definition}
\newtheorem{lemma}[definition]{Lemma}
\newtheorem{example}[definition]{Example}
\newcommand{\viz}{\dot{\sim}}

\firstsection{Introduction}

\maketitle

According to an English idiom, "A picture is worth a thousand words". Visual analytics aims to combine the power of visual perception with high performance computing in order to support human analytical reasoning. Since Pak Chung Wong and Jim Thomas published their article named "Visual Analytics" \cite{vis-an:wong2004visual} in 2004, visual analytics has been widely used in various fields such as biology or national security but also in other fields such as climate monitoring \cite{vis-an:climate:kollat2011many, vis-an:climate:scharl2013media} or social networks analysis, the field originally addressed by the Collaboration Spotting project. Networks built out of interconnected elements contained in datasets and represented as multidimensional, directed and labelled graphs are a natural means of representing data for visual analytics. 

Graphs as database models and graph query languages defined over these models have been investigated for some 30 years (Wood, P.T. \cite{Wood:2012:QLG:2206869.2206879}). These models and languages have been used in many applications using a wide spectrum of data (e.g. biology, social network and criminal investigation data), clearly indicating that the combination of visual analytics with graph query languages has become quite popular.
 
According to Wong, P.C. et al \cite{vis-an:challenge:wong2012top}, one of the biggest challenges in visual analytics is \textit{User-Driven Data Reduction} which calls for "\textit{a flexible mechanism that users can easily control according to their data collection practices and analytical needs}". This essentially entails an improvement of the visualization clarity and an escalation of data processing performances irrespective of the increasing complexity of the data over the years. To meet this challenge, semantic and structural abstraction techniques such as clustering, collapsing and extraction and demonstration of relationships among graph entities can be used  \cite{graph:growing:Davis:size_graph_exp} at the expanse of a loss of information on the network content \cite{shen2006visual}. 

The approach taken in the Collaboration Spotting project is to reduce the dimensional complexity of data networks while maintaining the information about their content. It consists in decomposing multi-dimensional, directed and labelled graphs  into multiple directed and weighted graphs of lesser dimensions - named views - and in building an interactive graph query language that supports user-specified views and full navigation across the data networks using these views as a support to the operations of the language. Within a view, structural abstraction techniques can then be used to enhance the visual perception further. This approach is quite similar to the concept of blueprint where the architectural plan is distributed across different views. The main advantage of this approach is that it combines \textit{Visual graph representation} and \textit{User interactions} \cite{Vis_an:survey:CGF:CGF1898} at the graph query language level.

Section \ref{section:related_work} gives a short overview on visualisation techniques for visual analytics (focusing on social networks) and on graph query languages fit to data networks. Section \ref{section:graph_def} gives a short description of the mathematical background supporting the approach. Section \ref{section:graph_creation}, introduces how views are constructed and Section \ref{graph:operations} shows how the basic operations of the query language enable users to conduct their analysis. In Section \ref{example}, the use-case that inspired the Collaboration Spotting project and the graphical query language are presented. This paper ends with conclusions and future work in Section \ref{conclusion_future}.

\section{Related Work}\label{section:related_work}
The related work is twofold since it combines multiple visual analytics techniques with the power of graph query languages. In the last 15 years, a lot of visual analytics articles were published with the aim of showing processes of transformation of multidimensional data into node-link diagrams \cite{Vis_an:survey:CGF:CGF1898,vis:survey:Multifaced}. 

A lot of articles have been published, especially on the \textit{coordinated multiple views} topic, which introduces a visual analytics paradigm supported by an interactive query language or by a set of operations. These articles can be divided into four different groups:
\begin{itemize}
\item OLAP \cite{software:olap:gray1997data} inspired paradigms that are using operations like \textit{slice, roll-up, dice}, etc. The most relevant papers are PivotGraph \cite{sotfware:pivot:Wattenberg:2006:VEM:1124772.1124891}, ScatterDice  \cite{software:scatterDice} (and GraphDice \cite{software:graphdice:Bezerianos:2010:GSE:2421836.2421849}), MatrixCube \cite{software:MatrixCube:Bach:2014:VDN:2556288.2557010} and Orion \cite{software:orion:heer2014orion}.

\item Relational algebra-related solutions such as Cross-filter views \cite{software:cross:5204083} which uses \textit{grouping, filtering, projection} and \textit{selection} operations, Polaris \cite{software:polaris:981851} that introduces and maps its algebra to SQL and Ploceus \cite{software:ploceus} which works with first-order logic language.
\item Other solutions such as Cross-filter views with hypergraph query language \cite{software:hypergraph:6634154}, JUNG\cite{software:jung} and Gephi\cite{software:gephi} that allow users to use other programming languages (JAVA in these cases).  
\end{itemize}

Literature on graph query languages is huge  \cite{graph_model:Hidders_goal,graph_model:Gyssens90agraph-oriented,graph_model:hidders2003typing,graph_model:guting1994graphdb,graph_model:Kunii:1987:DGD:42040.42071,graph_model:BarceloBaeza:2013:QGD:2463664.2465216,graph_model:paredaens1995g,graph_model:Yang:2011:DIQ:2063576.2063832}. It covers the use of different graph models reflecting the variety of requirements for applications and languages.

The visual analytics model introduced in this paper promotes a different approach to graph query language. The language operates on a directed, labelled graph that is managed via user interactions treated as query inputs and follows the semantic web query language concept, SPARQL \cite{harris2013sparql}. This approach allows users to generate graph patterns and evaluate them directly on the graph.

\section{Basic graph and views} \label{section:graph_def}
\begin{figure}[!tbh]
\centering
\includegraphics[page=3, height=2.5in]{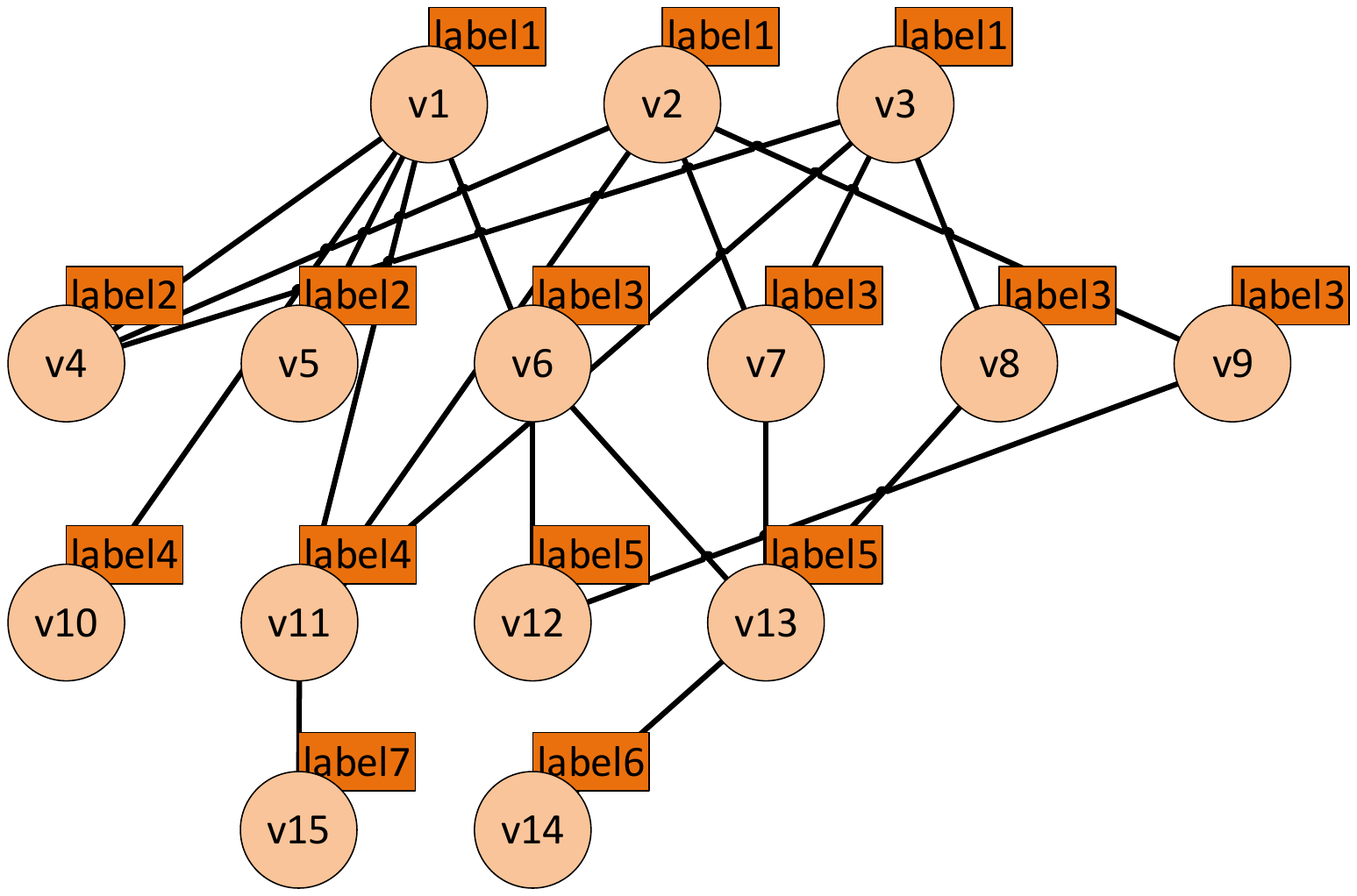}
\caption{Example of graph $G$ where $V = \{v1, \dots v15\}$ and $L = \{label1, \dots label7\}$} 
\label{figure:graph}
\end{figure}

Let graph $G$ be a directed, label   graph defined as a four-element tuple  $G=(V, E, L, \alpha)$ where $V$ represents a set of vertices and $E \subseteq V \times V$, a set of edges defined as a subset of the Cartesian products of these vertices. $L$ is a set of vertex labels and $\alpha: V \rightarrow L$ is a mapping function from vertices to the corresponding labels. Figure \ref{figure:graph} shows an example of such a graph. 
We define the reachability graph over graph $G$  as $G_{reachability}=(L, E_{reachability})$ where vertices are labels of graph $G$, $E_{reachability} \subseteq L \times L$ is defined as the Cartesian product of the labels where any two vertices of $G_{reachability}$ are connected if and only if there exists two connected vertices in graph $G$ and their respective labels correspond to the two vertices of graph $G_{reachability}$. Graph $G_{reachability}$ is a description of graph $G$, it is also called the graph schema of graph $G$. Graph schema helps users view graph $G$ via different sub-graphs of lesser dimensionality using labels of $G$ as dimensions and facilitates the generation of approximately optimal user-defined graph queries.
Let graph $G_{pattern} = (V_{pattern}, E_{pattern})$ be a graph pattern where $V_{pattern} \subseteq L$ and $E_{pattern} \subseteq E_{reachability} \cap V_{pattern} \times V_{pattern}$. To process the answer to a  graph query, one needs to find all possible isomorphic subgraphs of $G$ that are homomorphic to a graph pattern $G_{pattern}$ corresponding to the query.
This is a graph pattern matching problem, a well-known part of Mathematics \cite{gallagher2006matching}. In this case, one defines Graph $G' = (V', E', L, \alpha)$, a subgraph of graph $G$ as a sample matching the graph pattern $G_{pattern}$ if and only if:
\begin{itemize}
\item $\forall v' \in V' : \exists v \in V_{pattern},  \alpha(v') = v$,
\item $\forall (u', v') \in E' : (\alpha(u'), \alpha(v')) \in E_{pattern}$.
\end{itemize}  
The answer to a graph query is a view containing the set of subgraphs of $G$ matching $G_{pattern}$. To build such a view, one needs first to introduce the graph pairing function $pair$ and the set $Pattern$.  Let $G_{pattern_1}$ and $G_{pattern_2}$ be two graph patterns. These graph patterns are paired iff
\begin{itemize}
\item $V_{pattern_1} = V_{pattern_2}$ and
\item $\exists! a, b \in V_{pattern_1}: \textrm{ path(a, b)} \in E_{pattern_1} \textrm{ and path(a,b)} \not\in E_{pattern_2}, \\ \textrm{ path(b,a)} \not\in E_{pattern_1} \textrm{ and path(b, a)} \in E_{pattern_2}, E_{pattern_1} \setminus \textrm{ path(a,b)} = E_{pattern_2} \setminus \textrm{ path(b,a)}$.
\end{itemize}
Where a path is an alternate non-empty sequence of vertices and edges, starting and ending with vertices and requiring that all edges and vertices be distinct from one another. $\textrm{path(a,b)}\in E_{pattern_1}$ indicates that all edges of this path are in set $E_{pattern_1}$.
The $pair$ function is defined as 
\[
pair(G_{pattern}) := \left\{ 
\begin{array}{lc}
G_{pattern}^{pair}
& \textrm{if } G_{pattern}^{pair} \textrm{ is a pair of } G_{pattern} \\
(\emptyset, \emptyset) & \textrm{else.}
\end{array}
\right.
\]
And $Pattern$, the set of these pairs is defined as $Pattern := \{(g, g') | g,g' \textrm{are patterns}, g' = pair(g)\}$.

\begin{figure}[!tbh]
\centering
\includegraphics[page=7, height=2.5in]{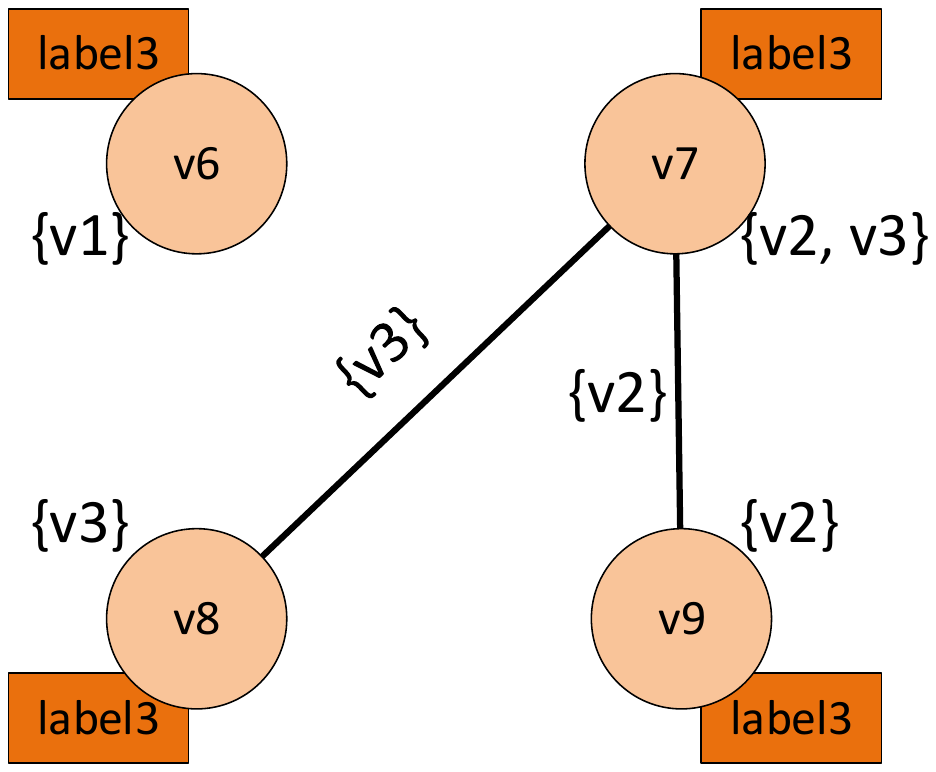}
\caption{Example of a view where $C_q = \{v6, \dots, v9\}, B_q = \{v1, \dots, v3\}, L_C = \{label3\}, L_B = \{label1\}$. The two graph patterns are $G_{pattern}= (\{label1, label3\}, \{(label1, label3)\})$ and pair($G_{pattern}^{rev}) = (\{label1, label3\}, \{(label3, label1)\})$} 
\label{figure:visualised_graph}
\end{figure}

A view of graph $G$ is defined as a six-element tuple $G_{q} = (C_q, B_q, E_q, L_q, \epsilon_q,  \upsilon_q,)$ where
\begin{itemize}
  \item $C_q \subset V, L_{C} := \{\alpha(v)| v \in C_q\}$, 
  \item $B_q \subset V, L_{B} := \{\alpha(b)| b \in B_q\}$, 
  \item $L_q \subseteq L$ and $L_q = L_{C} \cup L_{B}$,
  \item $E_q := \{(u,v)| u, v \in C_q, \exists G', G'' \subseteq G, \exists (G_{pattern},pair(G_{pattern})), (G_{pattern}^{rev},pair(G_{pattern}^{rev})) \in \textit{Patterns}: G' \textit{ matches } \textit{to } G_{pattern}, G'' \textit{ matches } \textit{to } \\pair(G_{pattern}^{rev}),  \exists b \in B_q: \textrm{path(}u, b \textrm{)} \in G', \textrm{path(}b, v \textrm{)} \in G''\}$,
  \item $\epsilon_q:E_q \rightarrow \mathcal{P}(B_q), \epsilon_q((u, v)) = \{b|b \in B_q, \exists G', G'' \subseteq G, \exists (G_{pattern},pair(G_{pattern})), (G_{pattern}^{rev},pair(G_{pattern}^{rev})) \in \textit{Patterns}: G' \textit{ matches }\textit{to } G_{pattern}, G'' \textit{ matches }\textit{to }\\ pair(G_{pattern}^{rev}), \textrm{path(}u, b \textrm{)} \in G', \textrm{path(}b, v \textrm{)} \in G''\}$,
  \item $\upsilon_q:C_q \rightarrow \mathcal{P}(B_q), \upsilon_q(u) = \{b|b \in B_q, \exists G' \subseteq G, \exists (G_{pattern},pair(G_{pattern})) \in \textit{Patterns}: G' \textit{ matches }\textit{to } G_{pattern}, \textrm{path(}u, b \textrm{)} \in G'\}$.
\end{itemize}
The use of multiple graph patterns for the construction of graph $G_q$ is required since the cardinality of set $L_B$ and set $L_C$ is not necessary equal to 1 (see details in Section \ref{section:graph_pattern}). To ease the reading, graph $G_q$ is noted $G_{L_B}^{L_C}$ to refer directly to the set of labels used in the construction of the view. Also, in practice, we use an aggregation function on edges, respectively on vertices in graph $G_q$ for determining their respective weights instead of the elements in set $B_q$ (for instance, the number of elements). Figure \ref{figure:visualised_graph} shows an example of a view. 

\section{Graph creation from user interactions} \label{section:graph_creation}
In this section, we introduce how the graph patterns and views can be created as a result of the following user interactions: 
\begin{itemize}
\item Selection of different nodes in the current view,
\item Removal of all vertices with the same label selected in one of the previous views,
\item Navigation from one view to another.
\end{itemize}
Users can modify set $L_C$ and set $L_B$ when performing any of the above interactions. Let $F \subseteq V$ be the set of vertices corresponding to a user selection, we define from $F$:
\begin{enumerate}
\item $L_F := \{l \in L| \exists f \in F: \alpha(f) = l\}$  which contains the labels of nodes in set $F$ and, 
\item $F_{|L^*} := \{f \in F| \alpha(f) \in L^* \}$ with $L^* \subseteq L$, a subset of set $F$, restricted to vertices having their respective labels in set $L^*$.
\end{enumerate}
In order for set $F$ to operate as a filter, the matched sample definition of Section \ref{section:graph_def} has to be restricted by requiring that $\forall v' \in V', \alpha(v') \in L_F \Rightarrow v' \in F$. Example \ref{def:filter} below shows the content of $L_F$ for user selection $F = \{v_4, v_6, v_7, v_{13}\}$ from the graph $G$ depicted in Figure \ref{figure:graph}.
\begin{example} \label{def:filter}
\begin{eqnarray}
F & = & \{v_4, v_6, v_7, v_{13}\} \label{example-f} \\
L_F & = & \{label_2, label_3, label_5 \} \label{example-fl}
\end{eqnarray}
\end{example}

\subsection{Graph pattern construction} \label{section:graph_pattern}
\begin{figure}[!tbh]
\centering
\includegraphics[page=16, height=2.5in]{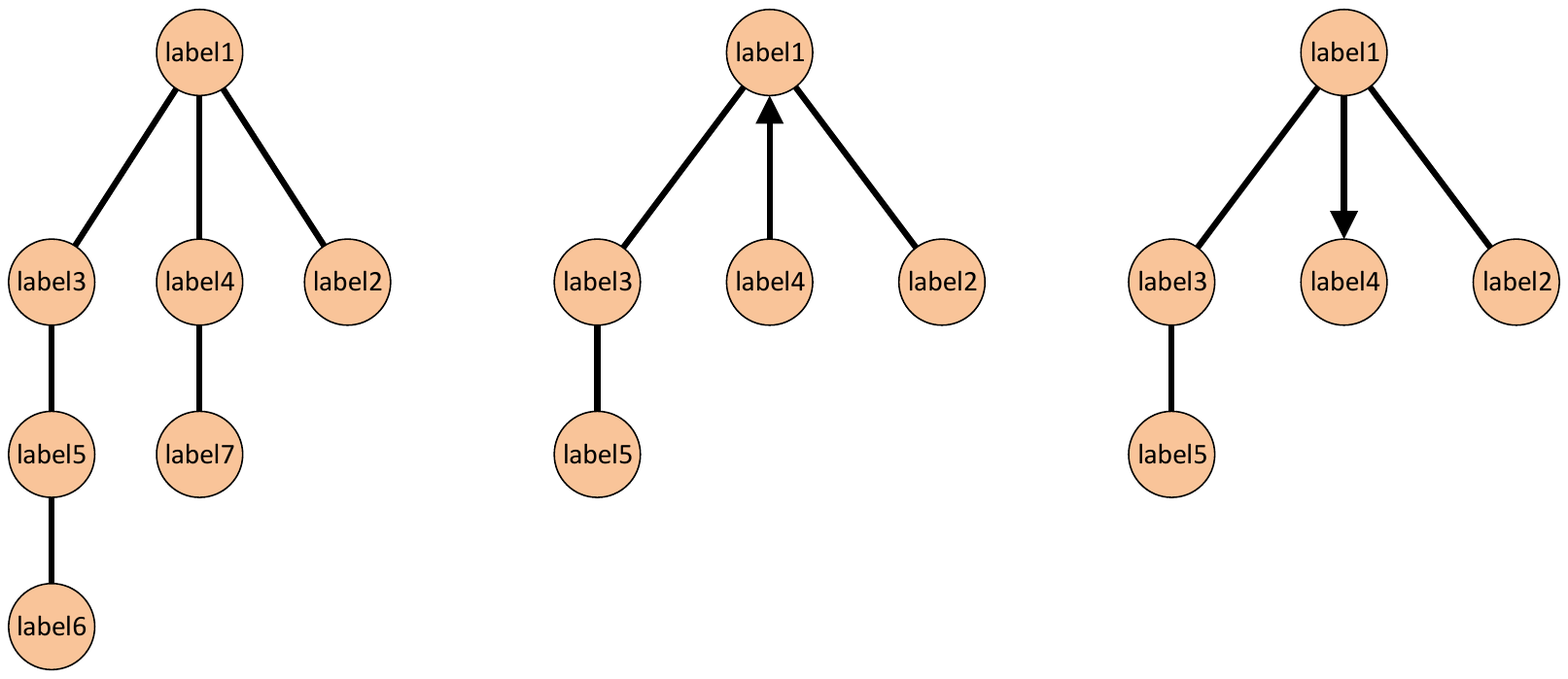}
\caption{On the left hand-side, the graph schema of graph $G$ in the example of Figure \ref{figure:graph}. On the middle and on the right hand-side, an example of a graph pattern pair for set F (Example \ref{example-f}), with $L_C = \{label4\}$ and $L_B = \{label1\}$}
\label{figure:schema_with_pattern}
\end{figure}

This section shows how to construct a graph pattern with set $L_F$ containing all the labels of vertices in set $F$.
We exploit the fact that graph patterns are actually only needed when constructing edges in $G_{L_B}^{L_C}$ and their respective weights. A pair of graph patterns are required for each combination of labels in set $L_C$ and set $L_B$ since the path direction between vertices from set $L_C$ and set $L_B$ are different due to the construction of edges between vertices of $C_q$ and vertices of $B_q$. Each pattern  has to satisfy the following criteria:
\begin{itemize}
\item It must be a connected and directed graph,
\item It must be minimal,
\item Labels from set $L \setminus L_F$ can be used as intermediate vertices in the pattern.
\end{itemize}
These requirements exactly fit a Steiner Minimal Tree problem \cite{hwang1992steiner}, known to be NP-complete\cite{garey1977complexity} and for which we use a minimal spanning tree solver as an approximation algorithm. 
Algorithm \ref{pattern_generator} describes the full process of pair generation. Figure \ref{figure:schema_with_pattern} shows the graph schema of graph $G$ depicted in Figure \ref{figure:graph} and the generated patterns.

\begin{algorithm}
\caption{Pattern generator algorithm}
\label{pattern_generator}
\begin{algorithmic}[1] 
\Function{PatternGenerator}{$F_L, L_B, L_C$}
\State $Patterns \gets \emptyset$
\State $B \gets L_B$
\While{$B \neq \emptyset$}
	\State $from, B \gets from \in B, B \setminus \{from\} $
	\State $E \gets L_C$
    \While{$E \neq \emptyset$}
    	\State $to, E \gets to \in E, E \setminus \{to\}$
        \State $Left \gets SpanningTree(F_L \cup \{from, to\}, from, to)$
        \State $Right \gets SpanningTree(F_L \cup \{from, to\}, to, from)$
        \State $Patterns \gets Patterns \cup \{(Left, Right)\}$
    \EndWhile
\EndWhile
\State \textbf{return} $Patterns$
\EndFunction
\end{algorithmic}
\end{algorithm}

\subsection{Connecting user interactions and views}
Now that graph patterns ($Patterns$) have been created using set $F$, set $L_C$ and set $L_B$, one can introduce the $gen$ function 
$gen: \mathcal{P}(V) \times \mathcal{P}(L) \times \mathcal{P}(L) \rightarrow G_{L_B}^{L_C}$ that generates views from user interactions. ($F \subseteq \mathcal{P}(V)$ and $L_C, L_B \subset \mathcal{P}(L)$) as \\ $gen(F, L_C, L_B) :=~^FG_{L_B}^{L_C} = (^FC_q, ^FB_q, E_q, L_q, \upsilon_q, \epsilon_q)$ where 
\[
^FC_q := \left\{ \begin{array}{cc} V \cap F_{|L_C} & \textrm{if } V \cap F_{|L_C} \neq \emptyset \\
V_{|L_C} & \textrm{else}
\end{array}\right.
\]
are the vertices of graph $^FG_{L_B}^{L_C}$ and
\[
^FB_q := \left\{ b \in {V'}_{|L_B}~ \left| 
\begin{array}{l}
\exists G' = (V', E', L, \alpha) \subseteq G,: \\ 
G' \textrm{ matches to } G_{pattern}, \forall v' \in V' : \\
(v' \in F \textrm{ or } \alpha(v') \not\in L_F)
\end{array}
\right. \right.
\]
are the "interconnection" vertices:
The other members of the six-tuple $G_q$ are unchanged since
\begin{itemize}
\item labels (set $L_q$) are not modified and since
\item edge definition (set $E_q$) and weighting functions ($\upsilon_q$ and $\epsilon_q$) only depend on set $^FC_q$ and set $^FB_q$.
\end{itemize}

\section{Operations on graphs} \label{graph:operations}
User interactions will result in the following graph operations:
\begin{itemize}
\item \textit{Selection}: The user selects nodes on the view,
\item \textit{Expansion}: The user expands a view by removing in his previous selection, vertices having the same labels,
\item \textit{Navigation}: The user navigates from a view to another. 
\end{itemize}
To define these operations one needs first to introduce the concepts of visual equivalence and minimal views since there can be views with vertices of null weight that are hidden to the user and hence non-selectable. Let $F_1$ and $F_2$ be two different filters on the same view complying with $F_1 \setminus F_{1|L_C} = F_2 \setminus F_{2|L_C}$. In essence, this means that there is no difference in the sets of vertices with labels contained in $L \setminus L_C$ which technically should be empty. View $^{F_1}G_{L_B}^{L_C}$ and view $^{F_2}G_{L_B}^{L_C}$
generated using F1 and F2 are said to be visual equivalent if and only if 
\begin{definition}[Vis-equivalent] \label{vis_eq}
\[
^{F_1}G_{L_B}^{L_C}~\viz~^{F_2}G_{L_B}^{L_C} \Leftrightarrow 
\begin{array}{l}
\forall v \in V_1 \setminus V_{2}: \upsilon_q(v) = \emptyset, \\
\forall v' \in V_{2} \setminus V_1: \upsilon_{q}(v') = \emptyset,
\end{array}
\]
\end{definition}
where $V_1$ ($V_2$) represents the vertices of view $^{F_1}G_{L_B}^{L_C}$ ($^{F_2}G_{L_B}^{L_C}$). 
Intuitively visual equivalence guaranties that vertices that are not common to two views have empty weights. It provides equivalence classification on views. It is easy to prove that for each class of views there is only one which does not have vertices with empty weights. This view is called the minimal view.

\subsection{Selection on graphs}
Let $F_{select}$ be the set of user selected nodes within a view. $F_{select} \subseteq V$ and $F_{select} \subseteq V'$ where $V'$ is a set of vertices from the minimal view which is visual-equivalent to graph $^FG^{L_C}_{L_B}$. The selection operator $\sigma : G_q \times \mathcal{P}(V) \rightarrow G_q$ is defined as 
\begin{definition}(Selection)
\[
\sigma(^FG^{L_C}_{L_B}, F_{select}) := \left\{ 
\begin{array}{l}
gen(F \cup F_{select}, L_C, L_B) \\
\qquad \qquad \textrm{if } F \cap F_{select} = \emptyset \\
gen((F \setminus F_{|L_C}) \cup F_{select}, L_C, L_B) \\ 
\qquad \qquad \textrm{else,} 
\end{array}
\right.
\]
\end{definition}
where $F_{|L_C} = \{f | f\in F, \alpha(f) \in L_C\}$. It is to be noted that at view creation the  the selection operator has been used with a more general definition of the $gen$ function.

\subsection{Expansion on graphs}
The expansion operator $\xi$ is in some sense the "invert" or the selection operator. It is defined as 
\begin{definition}(Expansion)
\[
\xi(^FG^{L_C}_{L_B}, L_{C'}) := gen(F \setminus F_{|L_{C'}}, L_{C'}, L_B).
\]
\end{definition}
The expansion operator changes view when $L_{C'} \neq L_C$ and remove all vertices in set $F$ that are labelled with labels in $L_C$.

\subsection{Navigation through graphs}
By selecting a subset of labels from $L_C$ one can build views of graph $G$ with reduced dimensional complexity. Navigation across views is required to enable users to apprehend the full graph $G$. Therefore the navigation function $\eta$ goes from view $^{F}G_{L_B}^{L_C}$ to a view labelled as $L_{C'}$ and $L_{B'}$ and is defined as:

\begin{definition}[Navigation]
\[
\eta (^FG^{L_C}_{L_B}, L_{C'}, L_{B'}) := gen(F, L_{C'}, L_{B'})
\]
\end{definition}

\subsection{Navigation history}
The navigation history can be represented as a navigation graph $G_{nav}$ where vertices represent navigation states and edges navigation steps between states. $G_{nav} = (N_{nav}, E_{nav})$ complies to
\begin{itemize}
\item $N_{nav} \subset \mathcal{P}(V) \times \mathcal{P}(L) \times \mathcal{P}(L)$. 
\item $E_{nav} \subseteq N_{nav} \times N_{nav} \times \{\sigma, \xi, \eta\}$,
\end{itemize}
where there is a navigation step between node $n_1 = (F_1, L_{C_1}, L_{B_1})$ to node $n_2 = (F_2, L_{C_2}, L_{B_2})$ if and only if one of the following statements is true:
\begin{enumerate}
\item $\sigma(gen(F_1, L_{C_1}, L_{B_1}), F_2 \setminus F_1) = gen(F_2, L_{C_2}, L_{B_2})$, and $L_{C_1} = L_{C_2}, L_{B_1} = L_{B_2}$; 
\item $\xi(gen(F_1, L_{C_1}, L_{B_1}), L_{C_2}) = gen(F_2, L_{C_2}, L_{B_2})$, and $F_2 = F_1 \setminus F_{1|L_{C_2}}, L_{B_1} = L_{B_2}$;
\item $\eta(gen(F_1, L_{C_1}, L_{B_1}), L_{C_2}, L_{B_2}) = gen(F_2, L_{C_2}, L_{B_2})$ and $F_1 = F_2$.
\end{enumerate}

In $E_{nav}$, the third component of an edge is always one of the operations $\sigma, \xi$ or $\eta$. It indicates how the step was processed. \\
The proper size of $N_{nav}$ is $2^n * (3^m - 2^{m+1} +1)$ where $n = |V|$ and $m = |L|$. \\
A particular navigation history corresponds to a walk in $G+{nav}$. An example of such a walk is given below 
\begin{example}[Walk on graph]\label{graph:walk}
\begin{eqnarray*}
(F_0, L_{C_0}, L_{B_0}),\eta, (F_1, L_{C_1}, L_{B_1}), \sigma, \dots, \xi,(F_f, L_{C_v}, L_{B_b})
\end{eqnarray*}
\end{example}

In practice, a particular set of labels $L_{C_0}$ is used to create an entry view from which all the above mentioned operations can then be performed by users.

\section{Use-case}\label{example}
In the framework of AIDA \cite{aida}, an FP7 project on Advanced European Infrastructures for Detectors at Accelerators, researchers needed to identify key players from academia and industry for technologies considered as strategic for the particle physics programme. To this end, the Collaboration Spotting project was launched in 2012 with a view to enabling users to search for technologies in titles and abstracts of publications and patents and viewing the organisation, journal category, keywords, city and country landscapes for each of these technologies individually. Individual technology searches are represented as vertices in a view named Technogram, used as the user entry view in which edges represent publications and/or patents common to searches. 

\subsection{Data}
Two different sources are used for searching. The metadata records of publications from Web of Science\texttrademark ~Core Collection \cite{thomson:Wok} developed by Clarivate Analytics (in the past, Thomson Reuters) and the metadata records of patents from PATSTAT developed by the European Patent Office \cite{epo:patstat}. Although the two sources have a number of labels in common, such as organisation, city and country there are others like journal category and keyword that only belong to publications. The subset of data from the two sources corresponding to the labels of interest for users was used to construct graph $G$ and its schema $G_{reachability}$.

\subsection{Storing data in a graph database (Neo4j)}
\begin{figure}[!tbh]
\centering
\includegraphics[page=2, height=2.5in]{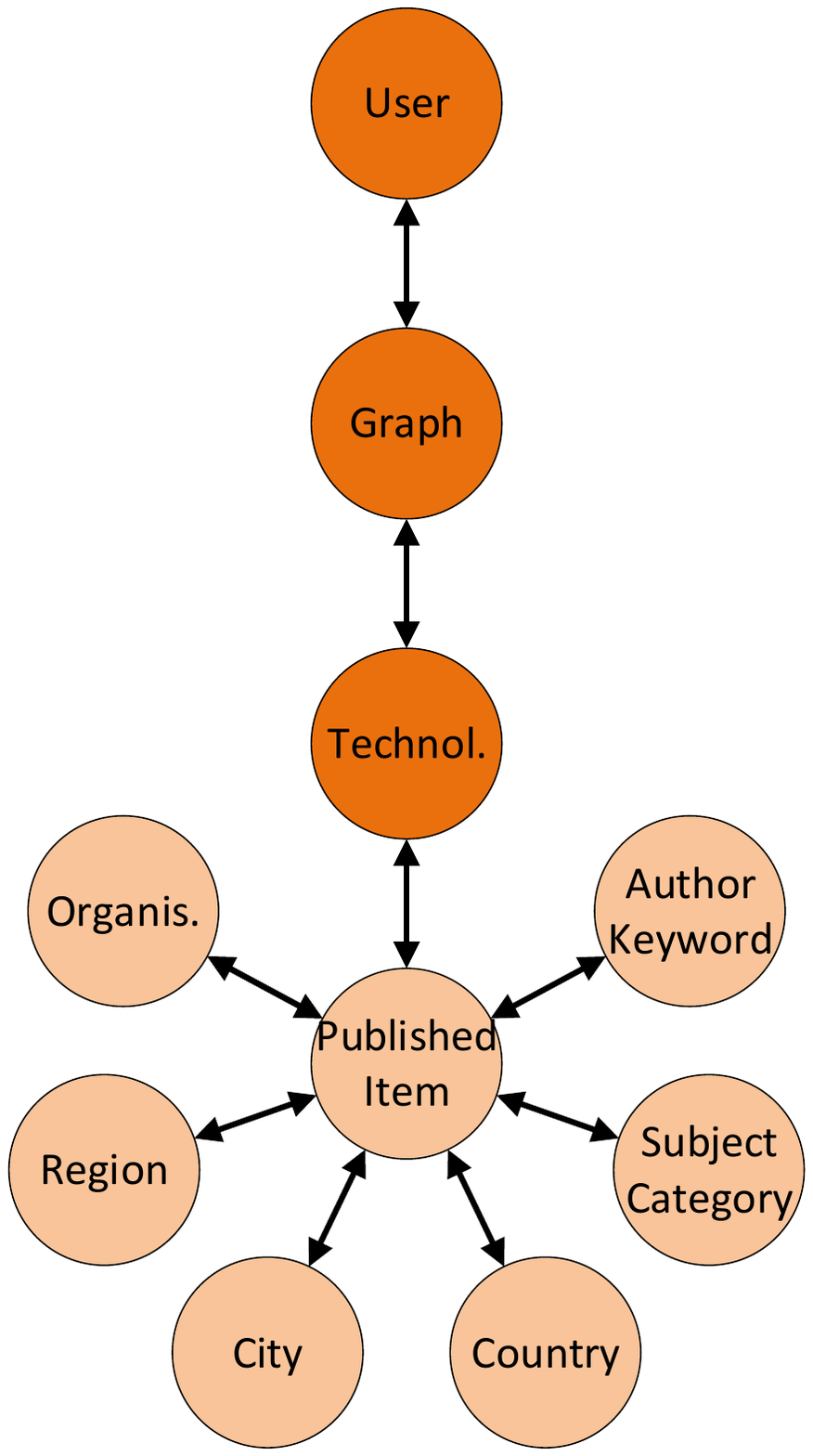}
\caption{The database schema (reachability graph); Light color nodes represent nodes uploaded by the data administrator and the dark nodes are created by the system itself by using search and authentication modules.}
\label{figure:schema}
\end{figure}

Graph $G$ is stored in a Neo4j graph database \cite{neo4j:manual}, in which individual metadata records are stored as subgraphs of labelled vertices using \textit{Published item, Organisation, Journal Category, Author Keyword, City, Region} and \textit{Country} as labels. Figure \ref{figure:schema} represents the reachability graph (graph schema) of this network.
Besides these labels, additional labels have been introduced to support user authentication and authorisation (\textit{User}) and searches (\textit{Graph} and \textit{Technology}). Searches use full text indices of the Apache Lucene project \cite{lucene} that have been integrated into the Neo4j database as legacy indices \cite{neo4j:manual}.

\subsubsection{Statistic of our graph data}
Searches on publications and patents metadata records from the 2000 - 2014 period can be performed. The resulting data network contains 45 million vertices and 150 million edges. Its breakdown is given in Table \ref{stat_on_nodes}. and Table \ref{stat_on_edges}.

\begin{table}[!tbh]
  \caption{Number of nodes by node labels}
  \begin{center} \label{stat_on_nodes}
      \begin{tabular}{| l | l |} \hline
      \textbf{Type of nodes} & \textbf{Number of nodes} \\ \hline
      Patents & 15.000.442 \\ \hline
      Publications & 20.087.904 \\ \hline
      Organisations & 2.918.060 \\ \hline
      Author Keywords & 8.193.604 \\ \hline
      Subject Categories & 230\\  \hline
      Cities & 7.741\\ \hline
      Regions & 946\\ \hline
      Countries & 128\\ \hline
      $\sum$ & 46.209.055\\ \hline
      \end{tabular}
  \end{center}
\end{table}

\begin{table}[!tbh]
  \caption{Number of edges by node labels. A patent does not have author keywords or subject categories property}
  \begin{center}\label{stat_on_edges}
      \begin{tabular}{l | l | l | l |}
      \cline{2-4}
      & \textbf{Patents} & \textbf{Publications} & \textbf{$\sum$}\\ 
      \cline{1-4} 
      \multicolumn{1}{|l|}{\textbf{Organ.}} & 12.440.903 & 36.672.677 &  49.113.580\\ \hline
      \multicolumn{1}{|l|}{\textbf{Author Key.}} & - & 48.941.098& 48.941.098\\ \hline
      \multicolumn{1}{|l|}{\textbf{Subject Cat.}} & - & 32.566.806 & 32.566.806\\ \hline
      \multicolumn{1}{|l|}{\textbf{Cities}} & 3.193.709 & 8.826.222 & 12.019.931\\ \hline
      \multicolumn{1}{|l|}{\textbf{Regions}} & 265.421 & 2.504.441 & 2.769.862\\ \hline
      \multicolumn{1}{|l|}{\textbf{Count.}} & 3.156.449 & 8.020.648 & 11.177.097\\ \hline
      \multicolumn{1}{|l|}{\textbf{$\sum$}} & 19.056.482& 137.531.892 & 156.588.374\\ \hline
      \end{tabular}
  \end{center}
\end{table}
As can be noticed the number of region edges is smaller than the number of country edges  due to the use of the $2^\textrm{nd}$ level of Nomenclature of Territorial Units For Statistic \cite{eu:nuts} created by the European Commission.

\subsection{Navigation}
The entry point for this use case is individual users. Using the terminology introduced above, the initial user interaction set $F$ contains user IDs.
\subsubsection{Limitations}
In the current implementation there is a restriction on the size of $L_C$ and $L_B$ fixed to a single label \textit{Published Item} and the visualization system only supports undirected edges. This calls for the generation of only one graph pattern instead of two making the system faster.

\begin{figure*}[!htbp]
 \centering
	\subfloat[Technology view ($L_C = \{\textit{Technology}\}, L_B = \{\textit{Published Item}\}$) of a user]{\includegraphics[page=17, width=0.47\textwidth]{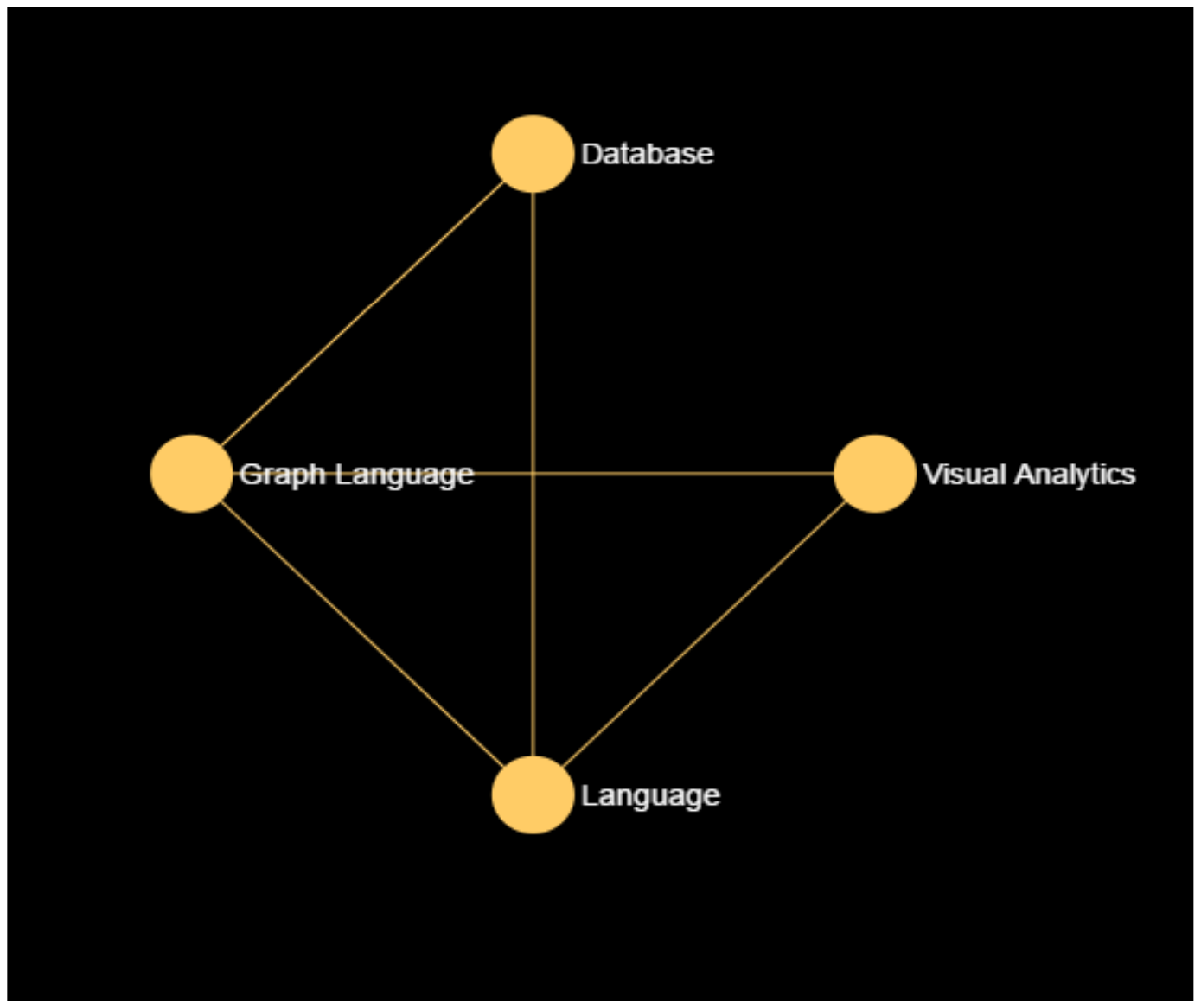}}
	\label{figure:technogram}
	\hfill
    \subfloat[Selecting two technologies ($F = \{\textit{Visual Analytics},\textit{Language}\}$) and navigating to $L_C = \{\textit{Subject Category}\}$ view.]{\includegraphics[page=18, width=0.47\textwidth]{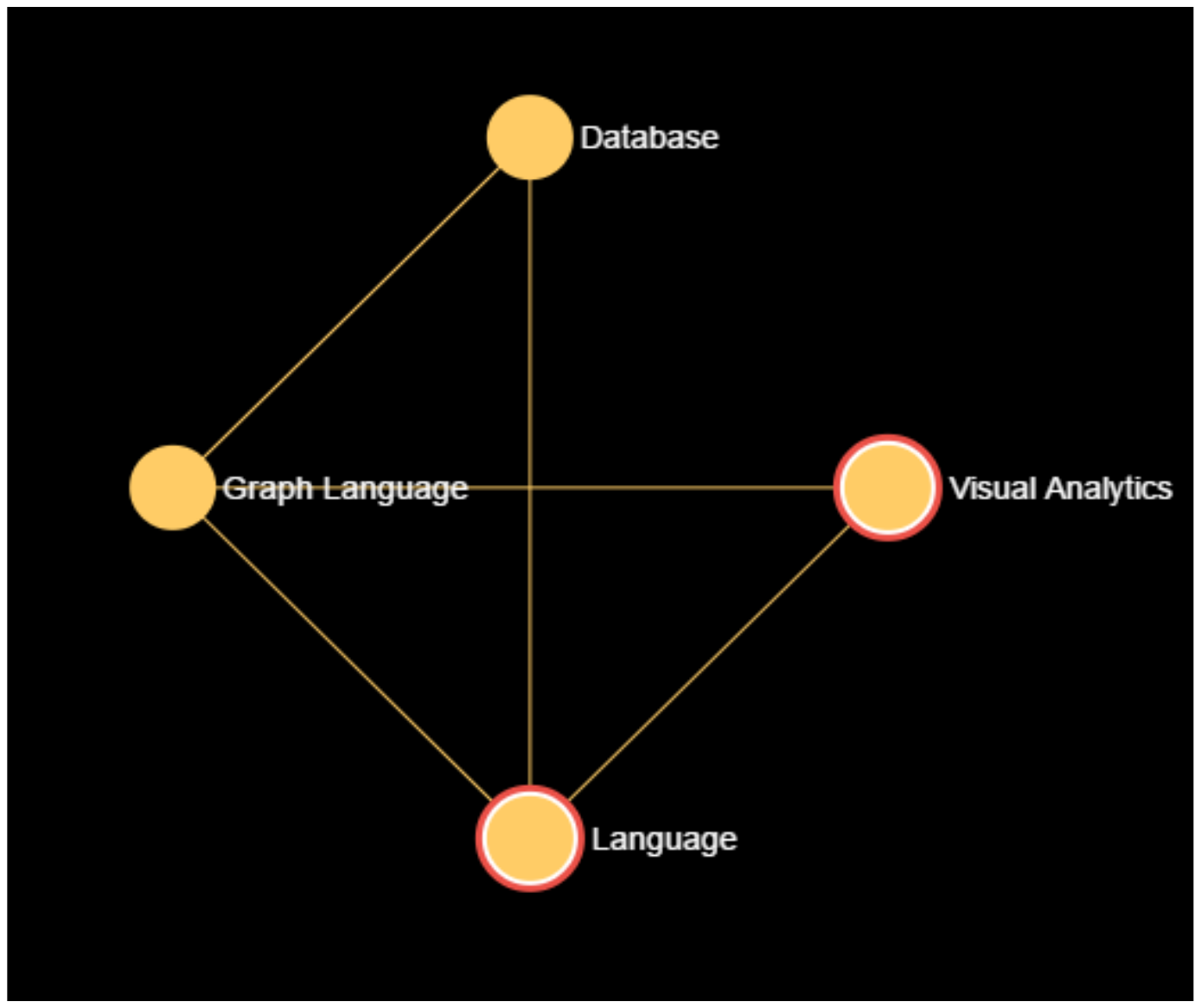}}
	\label{figure:select_and_navigate}
    \subfloat[Subject Category view ($L_C = \{\textit{Subject Category}\}$) for the selected technologies.]{\includegraphics[page=19, width=0.47\textwidth]{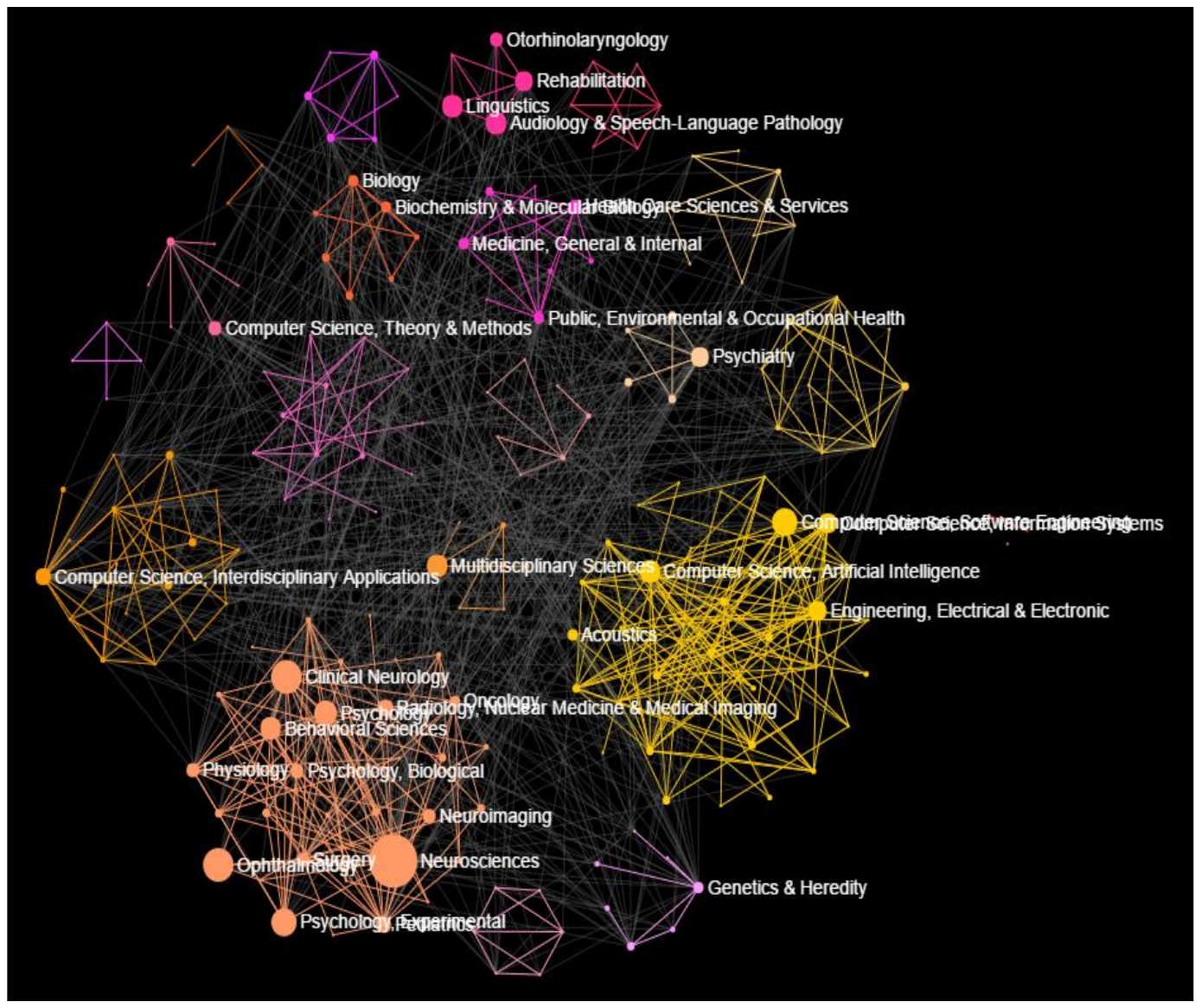}}
    \label{figure:go_to_subject_categorygram}
    \hfill
    \subfloat[Selecting a cluster in the Subject Category view ($F =\textit{Visual Analytics},\textit{Language, Lingustics} \dots \textit{Rehabilitation}\}$) and expanding the view to go back to the Technology view ($L_C = \{Technology\}$).]{\includegraphics[page=20, width=0.47\textwidth]{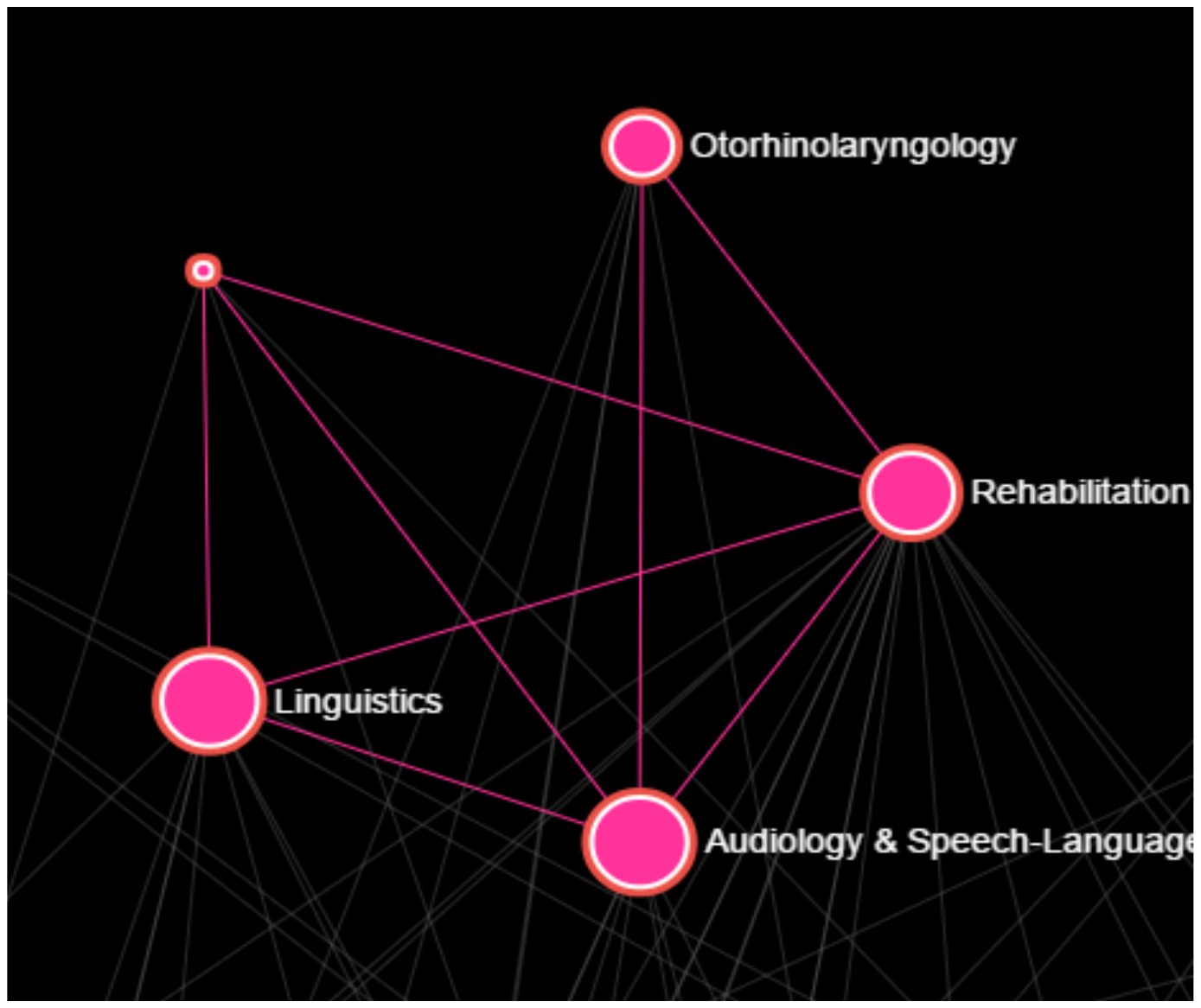}}
    \label{figure:expand_to}
    \subfloat[Technology view with $F = \textit{Language, Lingustics} \dots \textit{Rehabilitation}\}$ filter]{\includegraphics[page=21, width=0.47\textwidth]{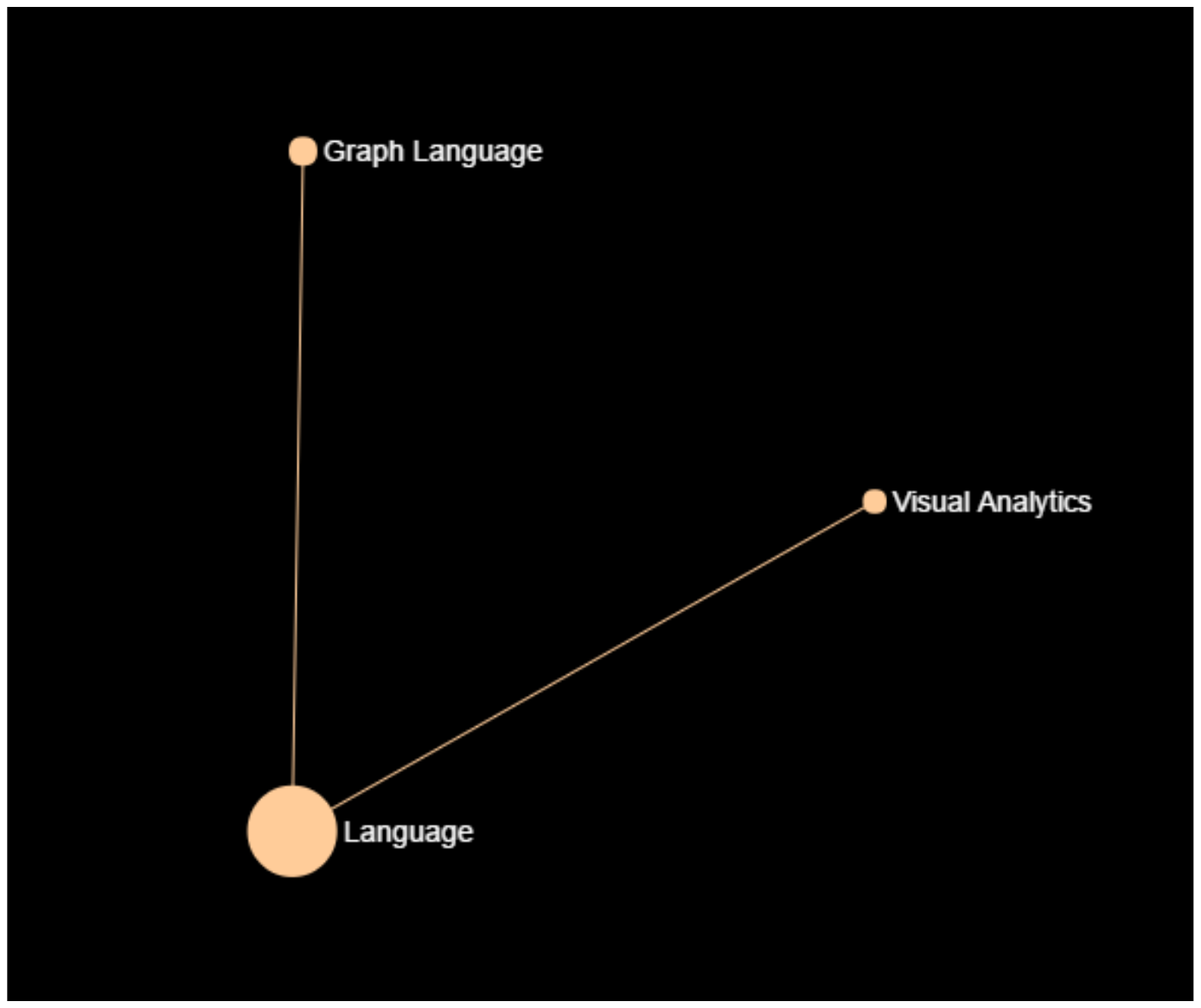}}
    \label{figure:end_of_demo}
	\caption{Example of operations; navigation, selection and expansion on views}
    \label{figure:example_for_operations}
\end{figure*}

In Figure \ref{figure:example_for_operations}, a short series of pictures illustrates how operations are working. The user enters the system with a technology view (vertices are labelled with the \textit{Technology} label and they are connected to the other views via vertices labelled with the \textit{Published Item} label).

\section{Conclusion and Future Work} \label{conclusion_future}
The current version of Collaboration Spotting running at CERN \cite{collspotting} addresses the implementation of the concepts using patents and publications metadata records. It is a new experimental service that aims to provide the High Energy Physics community (such as HEPTech \cite{heptech:website}) with information on Academia \& Industry main players active around key technologies, with a view to fostering more inter-disciplinary and inter-sectoral R\&D collaborations, and giving the procurement service the opportunity of reaching a wider selection of high-tech companies for biding purposes. Collaboration Spotting is generic in its concepts and implementation. It can support visual analytics of any kind of data and its backend is implemented using Neo4j graph database \cite{neo4j:manual}. Conference papers, technical \& business news, trademarks \& designs and financial data are amongst the data targeted to enrich the information on technologies that one can obtain from publications and patents. The choice of data sources will depend on users' priorities. The tool can be of use to other communities, in particular in dentistry\cite{dentistry:paper} but also to policy makers and investors if data in the labelled graph is enriched with technical \& business news and financial data.
Collaboration Spotting also addresses other types of data such as compatibility and dependency relationships in software and meta-data \cite{lhcb:cattaneo2012lhcb,lhcb:shapoval2014ariadne} of the LHCb experiment at CERN.

As an interactive graph query language, Collaboration Spotting is intended to provide a fully customisable visual analytics environment. In the current version data processing supports searches and contextual queries. In the future, labelled \& directed relationships and attributes on nodes will be included in the labelled property graph representation of the data network and the processing will be extended to more complex operations directly on the graph resulting from searches and queries with a view to enhancing the visual perception of users.

\acknowledgments{
}

\bibliographystyle{abbrv-doi}

\bibliography{template}

\begin{thebibliography}{10}

\bibitem{software:MatrixCube:Bach:2014:VDN:2556288.2557010}
B.~Bach, E.~Pietriga, and J.-D. Fekete.
\newblock Visualizing dynamic networks with matrix cubes.
\newblock In {\em Proceedings of the SIGCHI Conference on Human Factors in
  Computing Systems}, CHI '14, pp. 877--886. ACM, New York, NY, USA, 2014. doi:
  {{%
10\hspace{.1pt}\discretionary{.}{%
}{.}\hspace{.4pt}1145\discretionary{/}{%
}{/}2556288\hspace{.1pt}\discretionary{.}{%
}{.}\hspace{.4pt}2557010}}


\bibitem{graph_model:BarceloBaeza:2013:QGD:2463664.2465216}
P.~Barcel\'{o}~Baeza.
\newblock Querying graph databases.
\newblock In {\em Proceedings of the 32Nd Symposium on Principles of Database
  Systems}, PODS '13, pp. 175--188. ACM, New York, NY, USA, 2013. doi: {{%
10\hspace{.1pt}\discretionary{.}{%
}{.}\hspace{.4pt}1145\discretionary{/}{%
}{/}2463664\hspace{.1pt}\discretionary{.}{%
}{.}\hspace{.4pt}2465216}}


\bibitem{software:gephi}
M.~Bastian, S.~Heymann, M.~Jacomy, et~al.
\newblock Gephi: an open source software for exploring and manipulating
  networks.
\newblock {\em Icwsm}, 8:361--362, 2009.

\bibitem{software:graphdice:Bezerianos:2010:GSE:2421836.2421849}
A.~Bezerianos, F.~Chevalier, P.~Dragicevic, N.~Elmqvist, and J.~D. Fekete.
\newblock Graphdice: A system for exploring multivariate social networks.
\newblock In {\em Proceedings of the 12th Eurographics / IEEE - VGTC Conference
  on Visualization}, EuroVis'10, pp. 863--872. The Eurographs Association and
  John Wiley \& Sons, Ltd., Chichester, UK, 2010. doi: {{%
10\hspace{.1pt}\discretionary{.}{%
}{.}\hspace{.4pt}1111\discretionary{/}{%
}{/}j\hspace{.1pt}\discretionary{.}{%
}{.}\hspace{.4pt}1467\discretionary{%
}{-}{-}8659\hspace{.1pt}\discretionary{.}{%
}{.}\hspace{.4pt}2009\hspace{.1pt}\discretionary{.}{%
}{.}\hspace{.4pt}01687\hspace{.1pt}\discretionary{.}{%
}{.}\hspace{.4pt}x}}


\bibitem{lhcb:cattaneo2012lhcb}
M.~Cattaneo, M.~Clemencic, and I.~Shapoval.
\newblock {LHCb software and Conditions Database cross-compatibility tracking
  system: A graph-theory approach}.
\newblock In {\em Nuclear Science Symposium and Medical Imaging Conference
  (NSS/MIC), 2012 IEEE}, pp. 990--996. IEEE, 2012.

\bibitem{aida}
{CERN - AIDA team}.
\newblock {Advanced European Infrastructures for Detectors at Accelerators},
  December 2017.

\bibitem{thomson:Wok}
{Clarivate Analytics (in the past, Thomson Reuters)}.
\newblock {Web of Science\texttrademark}, December 2017.

\bibitem{collspotting}
{Collspotting Developer Team}.
\newblock {Collspotting}, December 2017.

\bibitem{graph:growing:Davis:size_graph_exp}
T.~A. Davis and Y.~Hu.
\newblock The university of florida sparse matrix collection.
\newblock {\em ACM Trans. Math. Softw.}, 38(1):1:1--1:25, Dec. 2011.

\bibitem{software:scatterDice}
N.~Elmqvist, P.~Dragicevic, and J.~D. Fekete.
\newblock Rolling the dice: Multidimensional visual exploration using
  scatterplot matrix navigation.
\newblock {\em IEEE Transactions on Visualization and Computer Graphics},
  14(6):1539--1148, Nov 2008. doi: {{%
10\hspace{.1pt}\discretionary{.}{%
}{.}\hspace{.4pt}1109\discretionary{/}{%
}{/}TVCG\hspace{.1pt}\discretionary{.}{%
}{.}\hspace{.4pt}2008\hspace{.1pt}\discretionary{.}{%
}{.}\hspace{.4pt}153}}


\bibitem{eu:nuts}
{European Commision}.
\newblock {NUTS - Nomenclature Of Territorial Units For Statistics}, December
  2017.

\bibitem{epo:patstat}
{European Patent Office}.
\newblock {PATSTAT - Worldwide Patent Statistical Database}, December 2017.

\bibitem{gallagher2006matching}
B.~Gallagher.
\newblock Matching structure and semantics: A survey on graph-based pattern
  matching.
\newblock {\em AAAI FS}, 6:45--53, 2006.

\bibitem{garey1977complexity}
M.~R. Garey, R.~L. Graham, and D.~S. Johnson.
\newblock The complexity of computing steiner minimal trees.
\newblock {\em SIAM journal on applied mathematics}, 32(4):835--859, 1977.

\bibitem{software:olap:gray1997data}
J.~Gray, S.~Chaudhuri, A.~Bosworth, A.~Layman, D.~Reichart, M.~Venkatrao,
  F.~Pellow, and H.~Pirahesh.
\newblock Data cube: A relational aggregation operator generalizing group-by,
  cross-tab, and sub-totals.
\newblock {\em Data mining and knowledge discovery}, 1(1):29--53, 1997.

\bibitem{graph_model:guting1994graphdb}
R.~H. G{\"u}ting.
\newblock Graphdb: Modeling and querying graphs in databases.
\newblock In {\em VLDB}, vol.~94, pp. 12--15. Citeseer, 1994.

\bibitem{graph_model:Gyssens90agraph-oriented}
M.~Gyssens, J.~Paredaens, J.~V.~D. Bussche, and D.~V. Gucht.
\newblock A graph-oriented object database model, 1990.

\bibitem{harris2013sparql}
S.~Harris, A.~Seaborne, and E.~Prud’hommeaux.
\newblock Sparql 1.1 query language.
\newblock {\em W3C Recommendation}, 21, 2013.

\bibitem{software:orion:heer2014orion}
J.~Heer and A.~Perer.
\newblock Orion: A system for modeling, transformation and visualization of
  multidimensional heterogeneous networks.
\newblock {\em Information Visualization}, 13(2):111--133, 2014.

\bibitem{heptech:website}
{HEPTech Team}.
\newblock {HEPTech - website}, December 2017.

\bibitem{graph_model:hidders2003typing}
J.~Hidders.
\newblock Typing graph-manipulation operations.
\newblock In {\em Database Theory-ICDT 2003}, pp. 394--409. Springer, 2003.

\bibitem{graph_model:Hidders_goal}
J.~Hidders and J.~Paredaens.
\newblock {GOAL, A Graph-based Object and Association Language}.

\bibitem{hwang1992steiner}
F.~K. Hwang, D.~S. Richards, and P.~Winter.
\newblock {\em The Steiner tree problem}, vol.~53 of {\em Annals of Discrete
  Mathematics}.
\newblock Elsevier, 1992.

\bibitem{vis:survey:Multifaced}
J.~Kehrer and H.~Hauser.
\newblock Visualization and visual analysis of multifaceted scientific data: A
  survey.
\newblock {\em IEEE Transactions on Visualization and Computer Graphics},
  19(3):495--513, March 2013. doi: {{%
10\hspace{.1pt}\discretionary{.}{%
}{.}\hspace{.4pt}1109\discretionary{/}{%
}{/}TVCG\hspace{.1pt}\discretionary{.}{%
}{.}\hspace{.4pt}2012\hspace{.1pt}\discretionary{.}{%
}{.}\hspace{.4pt}110}}


\bibitem{vis-an:climate:kollat2011many}
J.~B. Kollat, P.~M. Reed, and R.~M. Maxwell.
\newblock Many-objective groundwater monitoring network design using bias-aware
  ensemble kalman filtering, evolutionary optimization, and visual analytics.
\newblock {\em Water Resources Research}, 47(2), 2011.
\newblock W02529.

\bibitem{graph_model:Kunii:1987:DGD:42040.42071}
H.~S. Kunii.
\newblock {DBMS with Graph Data Model for Knowledge Handling}.
\newblock In {\em Proceedings of the 1987 Fall Joint Computer Conference on
  Exploring Technology: Today and Tomorrow}, ACM '87, pp. 138--142. IEEE
  Computer Society Press, Los Alamitos, CA, USA, 1987.

\bibitem{dentistry:paper}
E.~Leonardi, A.~Agocs, S.~Fragkiskos, N.~Kasfikis, J.~Le~Goff, M.~Cristalli,
  V.~Luzzi, and A.~Polimeni.
\newblock Collaboration spotting for dental science.
\newblock {\em Minerva Stomatologica}, 63(9):295--306, sep 2014.

\bibitem{software:ploceus}
Z.~Liu, S.~B. Navathe, and J.~T. Stasko.
\newblock Network-based visual analysis of tabular data.
\newblock In {\em 2011 IEEE Conference on Visual Analytics Science and
  Technology (VAST)}, pp. 41--50, Oct 2011. doi: {{%
10\hspace{.1pt}\discretionary{.}{%
}{.}\hspace{.4pt}1109\discretionary{/}{%
}{/}VAST\hspace{.1pt}\discretionary{.}{%
}{.}\hspace{.4pt}2011\hspace{.1pt}\discretionary{.}{%
}{.}\hspace{.4pt}6102440}}


\bibitem{lucene}
{Lucene\texttrademark/Solr\texttrademark~Committers}.
\newblock {\em {Apache Lucene\texttrademark~Documentation}}, December 2017.

\bibitem{neo4j:manual}
Neo4j.
\newblock {\em The Neo4j Manual v2.3.3}, December 2017.

\bibitem{software:jung}
J.~O'Madadhain, D.~Fisher, S.~White, and Y.~Boey.
\newblock {The JUNG (Java Universal Network/Graph) Framework}.
\newblock {\em University of California, Irvine, California}, 2003.

\bibitem{graph_model:paredaens1995g}
J.~Paredaens, P.~Peelman, and L.~Tanca.
\newblock {G-Log: A graph-based query language}.
\newblock {\em Knowledge and Data Engineering, IEEE Transactions on},
  7(3):436--453, 1995.

\bibitem{vis-an:climate:scharl2013media}
A.~Scharl, A.~Hubmann-Haidvogel, A.~Weichselbraun, H.~P. Lang, and M.~Sabou.
\newblock Media watch on climate change -- visual analytics for aggregating and
  managing environmental knowledge from online sources.
\newblock In {\em 2013 46th Hawaii International Conference on System
  Sciences}, pp. 955--964, Jan 2013. doi: {{%
10\hspace{.1pt}\discretionary{.}{%
}{.}\hspace{.4pt}1109\discretionary{/}{%
}{/}HICSS\hspace{.1pt}\discretionary{.}{%
}{.}\hspace{.4pt}2013\hspace{.1pt}\discretionary{.}{%
}{.}\hspace{.4pt}398}}


\bibitem{software:hypergraph:6634154}
R.~Shadoan and C.~Weaver.
\newblock Visual analysis of higher-order conjunctive relationships in
  multidimensional data using a hypergraph query system.
\newblock {\em IEEE Transactions on Visualization and Computer Graphics},
  19(12):2070--2079, Dec 2013. doi: {{%
10\hspace{.1pt}\discretionary{.}{%
}{.}\hspace{.4pt}1109\discretionary{/}{%
}{/}TVCG\hspace{.1pt}\discretionary{.}{%
}{.}\hspace{.4pt}2013\hspace{.1pt}\discretionary{.}{%
}{.}\hspace{.4pt}220}}


\bibitem{lhcb:shapoval2014ariadne}
I.~Shapoval, M.~Clemencic, and M.~Cattaneo.
\newblock {ARIADNE: a Tracking System for Relationships in LHCb Metadata}.
\newblock In {\em Journal of Physics: Conference Series}, vol. 513, p. 042039.
  IOP Publishing, 2014.

\bibitem{shen2006visual}
Z.~Shen, K.-L. Ma, and T.~Eliassi-Rad.
\newblock Visual analysis of large heterogeneous social networks by semantic
  and structural abstraction.
\newblock {\em IEEE transactions on visualization and computer graphics},
  12(6):1427--1439, 2006.

\bibitem{software:polaris:981851}
C.~Stolte, D.~Tang, and P.~Hanrahan.
\newblock Polaris: a system for query, analysis, and visualization of
  multidimensional relational databases.
\newblock {\em IEEE Transactions on Visualization and Computer Graphics},
  8(1):52--65, Jan 2002. doi: {{%
10\hspace{.1pt}\discretionary{.}{%
}{.}\hspace{.4pt}1109\discretionary{/}{%
}{/}2945\hspace{.1pt}\discretionary{.}{%
}{.}\hspace{.4pt}981851}}


\bibitem{Vis_an:survey:CGF:CGF1898}
T.~von Landesberger, A.~Kuijper, T.~Schreck, J.~Kohlhammer, J.~van Wijk, J.-D.
  Fekete, and D.~Fellner.
\newblock Visual analysis of large graphs: State-of-the-art and future research
  challenges.
\newblock {\em Computer Graphics Forum}, 30(6):1719--1749, 2011.

\bibitem{sotfware:pivot:Wattenberg:2006:VEM:1124772.1124891}
M.~Wattenberg.
\newblock Visual exploration of multivariate graphs.
\newblock In {\em Proceedings of the SIGCHI Conference on Human Factors in
  Computing Systems}, CHI '06, pp. 811--819. ACM, New York, NY, USA, 2006. doi:
  {{%
10\hspace{.1pt}\discretionary{.}{%
}{.}\hspace{.4pt}1145\discretionary{/}{%
}{/}1124772\hspace{.1pt}\discretionary{.}{%
}{.}\hspace{.4pt}1124891}}


\bibitem{software:cross:5204083}
C.~Weaver.
\newblock Cross-filtered views for multidimensional visual analysis.
\newblock {\em IEEE Transactions on Visualization and Computer Graphics},
  16(2):192--204, March 2010. doi: {{%
10\hspace{.1pt}\discretionary{.}{%
}{.}\hspace{.4pt}1109\discretionary{/}{%
}{/}TVCG\hspace{.1pt}\discretionary{.}{%
}{.}\hspace{.4pt}2009\hspace{.1pt}\discretionary{.}{%
}{.}\hspace{.4pt}94}}


\bibitem{vis-an:challenge:wong2012top}
P.~C. Wong, H.-W. Shen, C.~R. Johnson, C.~Chen, and R.~B. Ross.
\newblock The top 10 challenges in extreme-scale visual analytics.
\newblock {\em IEEE computer graphics and applications}, 32(4):63, 2012.

\bibitem{vis-an:wong2004visual}
P.~C. Wong and J.~Thomas.
\newblock Visual analytics.
\newblock {\em IEEE Computer Graphics and Applications}, 24(5):20--21, Sept
  2004. doi: {{%
10\hspace{.1pt}\discretionary{.}{%
}{.}\hspace{.4pt}1109\discretionary{/}{%
}{/}MCG\hspace{.1pt}\discretionary{.}{%
}{.}\hspace{.4pt}2004\hspace{.1pt}\discretionary{.}{%
}{.}\hspace{.4pt}39}}


\bibitem{Wood:2012:QLG:2206869.2206879}
P.~T. Wood.
\newblock Query languages for graph databases.
\newblock {\em SIGMOD Rec.}, 41(1):50--60, Apr. 2012.

\bibitem{graph_model:Yang:2011:DIQ:2063576.2063832}
J.~Yang, S.~Zhang, and W.~Jin.
\newblock {DELTA: Indexing and Querying Multi-labeled Graphs}.
\newblock In {\em Proceedings of the 20th ACM International Conference on
  Information and Knowledge Management}, CIKM '11, pp. 1765--1774. ACM, New
  York, NY, USA, 2011. doi: {{%
10\hspace{.1pt}\discretionary{.}{%
}{.}\hspace{.4pt}1145\discretionary{/}{%
}{/}2063576\hspace{.1pt}\discretionary{.}{%
}{.}\hspace{.4pt}2063832}}


\end{thebibliography}
\end{document}